# Half-kilowatt high energy third harmonic conversion to 50 J @ 10 Hz at 343 nm

Jan Pilar [1], Martin Divoky [1], Jonathan Phillips [2], Martin Hanus [1], Petr Navratil [1], Ondrej Denk [1], Patricie Severova [1], Tomas Paliesek [1], Danielle Clarke [2], Martin Smrz [1], Thomas Butcher [2], Chris Edwards [2] and Tomas Mocek [1]

[1] *HiLASE Centre, Institute of Physics of the Czech Academy of Sciences, Za Radnici 828, 25241, Dolni Brezany, Czech Republic*
[2] *Central Laser Facility, STFC Rutherford Appleton Laboratory, Didcot, OX11 0QX, UK*

**Abstract** We present results of frequency tripling experiments performed at Hilase facility on a cryogenically gas cooled multi-slab Yb:YAG laser system Bivoj/DiPOLE. The laser produces high energy ns pulses at 10 Hz repetition rate, which are frequency doubled using a type-I phase-matched LBO crystal and consequently frequency summed using a type-II phase-matched LBO crystal. We demonstrated a stable frequency conversion to 343 nm at 50 J energy and 10 Hz repetition rate with conversion efficiency of 53%.

## I. INTRODUCTION

With the rapid development of high energy and high average power lasers (HE-HAP) diode pumped laser systems, harmonic frequency conversion is a tool to broaden application spectrum of such lasers. Recent advances with indirectly driven thermonuclear fusion bring fusion power plant closer to our reach [1], which will require UV lasers with high repetition rate. Such lasers will need optical components LIDT tested with large aperture beam [2–3]. Other applications include annealing of Silicone to improve its electrical properties for semiconductor industry [4]. And with ultra short pulses, UV micromachining [5], UV ablation [5] and UV air ionization [5] offer performance increase over IR laser pulses.

Phillips et al. demonstrated 65 W of average power with energy of 65 J at 343 nm using multi slab laser system DiPOLE100 and harmonic conversion in lithium triborate (LBO) [6]. Rothhardt at al. demonstrated 100 W of average power and pulse energy of 28.5 uJ at 343 nm with femtosecond fiber laser and subsequent harmonic conversion in beta barium borate (BBO) crystals [7]. Andral et al. generated 120 W of average power and pulse energy of 118 mJ using thin disk picosecond laser and harmonic conversion in LBO crystals [8]. Negel et al. reported on third harmonic generation in LBO with average power output of 234 W at 343 nm in 7.7 ps pulses from a thin-disk laser operating at 300 kHz [9].

In this paper we report on Third Harmonic Generation (THG) of 55 J at 10 Hz repetition rate using output of Bivoj/DiPOLE laser system, which corresponds to the average power of 550 W and more than 2 times increase to the published state-of-the-art. We reached a conversion efficiency of 60% by using square flat top beam profile with top-hat pulse profile and by efficient optimization of thermally induced polarization changes by a previously developed polarimetric method described in [10]. After thermal stabilization, the third harmonic energy dropped to around 50 J at 10 Hz. Further increase of the energy at 343 nm was limited by energy available at the fundamental wavelength.

## II. EXPERIMENT

The setup for conversion to 515 nm was described in the previous paper [11] and used a sealed mount for temperature stabilization of the SHG LBO crystal, which resulted in better conversion stability. The frequency doubled output at 515 nm was converted to 343 nm using residual light at fundamental wavelength of 1030 nm and type II phase matching in second LBO crystal (Coherent Inc.) placed after the first LBO crystal. The crystal had an aperture of $60 \times 60$ mm$^2$, thickness of 12 mm and cut angles $\theta$ of 50.4° and $\varphi$ of 90°. Dual Band Anti-Reflection (DBAR) coating for 1030 nm and 515 nm was used on the front face to minimize reflection to 0.1% and 0.15%, respectively. Single Layer Anti-Reflection (SLAR) coating for 343 nm was used on the back face to minimize reflection to less than 1.5% at 343 nm. Reflection values of the SLAR coating at 1030 nm and 515 nm were 1% and 2.5%, respectively. The above mentioned values of reflectivity are provided by coatings manufacturer. The THG LBO crystal was placed in a mount provided with temperature stabilization (IB Photonics). Both LBO crystals were kept at 30°C as it provided the highest conversion efficiency. While the SHG oven was able to provide some cooling power through its thermoelectric stabilization unit, the THG oven was using only natural convection of ambient air as cooling mechanism.

The layout of the harmonic conversion setup is shown in Figure 1 and was described in our previous paper [12]. The output from Bivoj laser system is transferred to the harmonic conversion setup by using two Keplerian telescopes. To increase the energy fluence on the crystal, the telescope de-magnifies the laser beam size 1.56 times from $77 \times 77$ mm$^2$ to $49 \times 49$ mm$^2$.

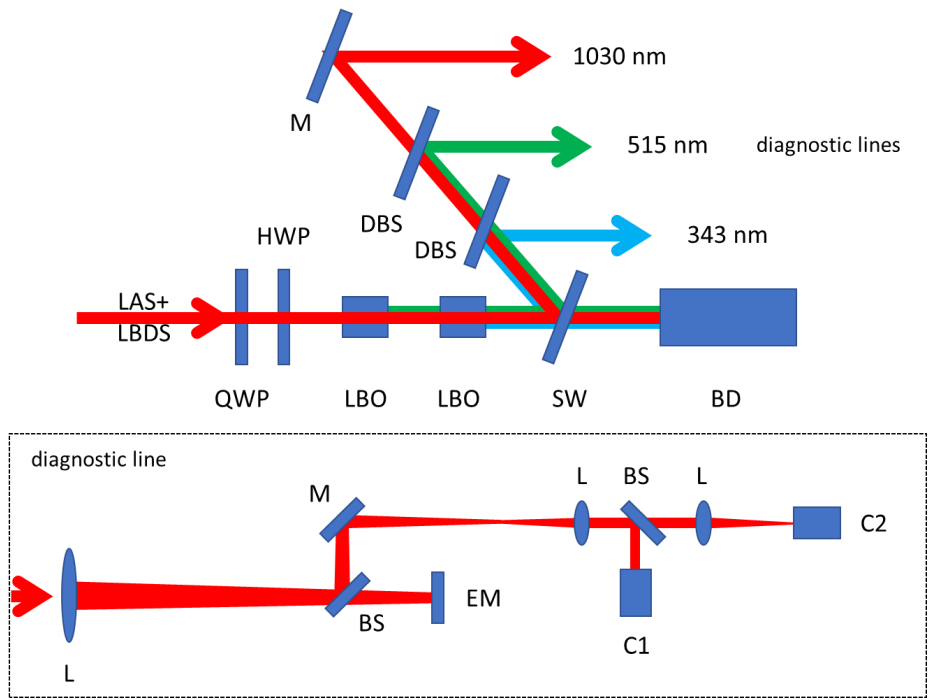

**Figure 1** Schematic layout of the conversion experiment. The laser beam is coming from the laser system via laser beam distribution system (LAS+LBDS). Single components of the setup are denoted as: quarter waveplate (QWP), half waveplate (HWP), conversion crystals (LBO), partially reflecting sampling wedge (SW) and beam dump (BD). Diagnostics consists of dichroic beamsplitter (DBS), mirrors (M), lenses (L), beamsplitters (BS), energy meter (EM), near field camera (C1) and far field camera (C2 – not present during the experiment). The layout of diagnostic lines is the same for all three wavelengths and is shown only once.

Two pairs of zero-order waveplates were used to compensate thermally induced polarization changes of the Bivoj final amplifier and the latter pair was used also to adjust polarization at the input of the LBO crystals in order to orient it parallel with the principal plane of the crystal and thus maximizing the overall conversion efficiency. After the conversion crystals, the beams at 1030 nm, 515 nm and 343 nm wavelengths were absorbed in a beam dump realized by a set of colored glass filters suspended in a water tank. In order to obtain a low power sample beam for diagnostics a dielectrically coated sampling wedge was placed in front of the beam dump with reflection values being close to 1% for 1030 nm, 515 nm and 343 nm wavelengths. The single components of the diagnostic beam were separated according to their wavelengths by a pair of dichroic mirrors. Each diagnostic line consisted of an energy meter (Gentec-EO QE25LP) and a diagnostic camera (AVT Manta 145B) to monitor image-relayed beam profile.

The energy meters were calibrated so that they would show the energy values present at the sampler for any of the three wavelengths. The calibration coefficients for the input (1030 nm), the unconverted fundamental (1030 nm) and second harmonic (515 nm) energy meters remained constant during the whole experiment. However, the calibration coefficient for the third harmonic frequency (343 nm) varied by more than 20% throughout the experiment. Evolution of calibration coefficients is shown in Figure 2. We suspect the cause was the change of reflectivity of the sampling wedge due to heating by the UV laser light. Because of the 3ω energy meter coefficient uncertainty, we decided to estimate the third harmonic energy as the difference between the input energy and the sum of energies of unconverted fundamental and second harmonic frequencies. The calculation of the 3ω energy makes use of all known values of transmissivity of all relevant optical coatings. The calculation estimates the 3ω energy at the end of THG LBO (before it enters the final coating) and applies the known value of coating transmission at 3ω. The values of transmissivity used in the calculation are listed in (1) below.

$$T_{SHG-1\omega} = 99.0\% \quad T_{in-THG-1\omega} = 99.5\% \quad T_{out-THG-1\omega} = 99.5\%$$
$$T_{SHG-2\omega} = 99.0\% \quad T_{in-THG-2\omega} = 99.5\% \quad T_{out-THG-2\omega} = 97.0\% \quad (1)$$
$$T_{out-THG-3\omega} = 98.0\%$$

Where the $T_{SHG-x\omega}$ denotes transmission through the whole SHG oven (2 windows and LBO crystal) for the given harmonics, $T_{in-THG-x\omega}$ denotes transmission through the first coating of THG LBO and $T_{out-THG-x\omega}$ denotes transmission through the second coating of the THG LBO for the given harmonics.

The input 1030 nm energy was decreased by transmission through the SHG oven and first optical surface of THG LBO crystal. (*98.5%*). The measured residual energies ($E_{1\omega}$, $E_{2\omega}$) were increased by dividing them by transmission of the second optical surface of THG LBO crystal. These energies were subtracted from the input to obtain 3ω energy at the THG LBO end and then decreased by the second optical surface transmission at 343 nm. The calculation is shown also in (2) below. Scattering and absorption losses were neglected as they should be an order of magnitude lower than losses on AR coatings.

$$E_{3\omega} = T_{out-THG-3\omega} \cdot \left(E_{input} \cdot T_{SHG} \cdot T_{in-THG} - \frac{E_{1\omega}}{T_{out-THG-1\omega}} - \frac{E_{2\omega}}{T_{out-THG-2\omega}}\right) \quad (2)$$

Where $E_{x\omega}$ denotes measured energy of the given harmonics, $T_{SHG-x\omega}$ denotes transmission through the whole SHG oven (2 windows and LBO crystal) for the given harmonics, $T_{in-THG-x\omega}$ denotes transmission through the first coating of THG LBO and $T_{out-THG-x\omega}$ denotes transmission through the second coating of the THG LBO for the given harmonics.

We plan to investigate this phenomenon further in a follow-up work as well as comparing the performance of our in-situ diagnostics to a measurement of diagnostic beam reflected from an uncoated wedge, which should not suffer from thermally induced reflection variations.

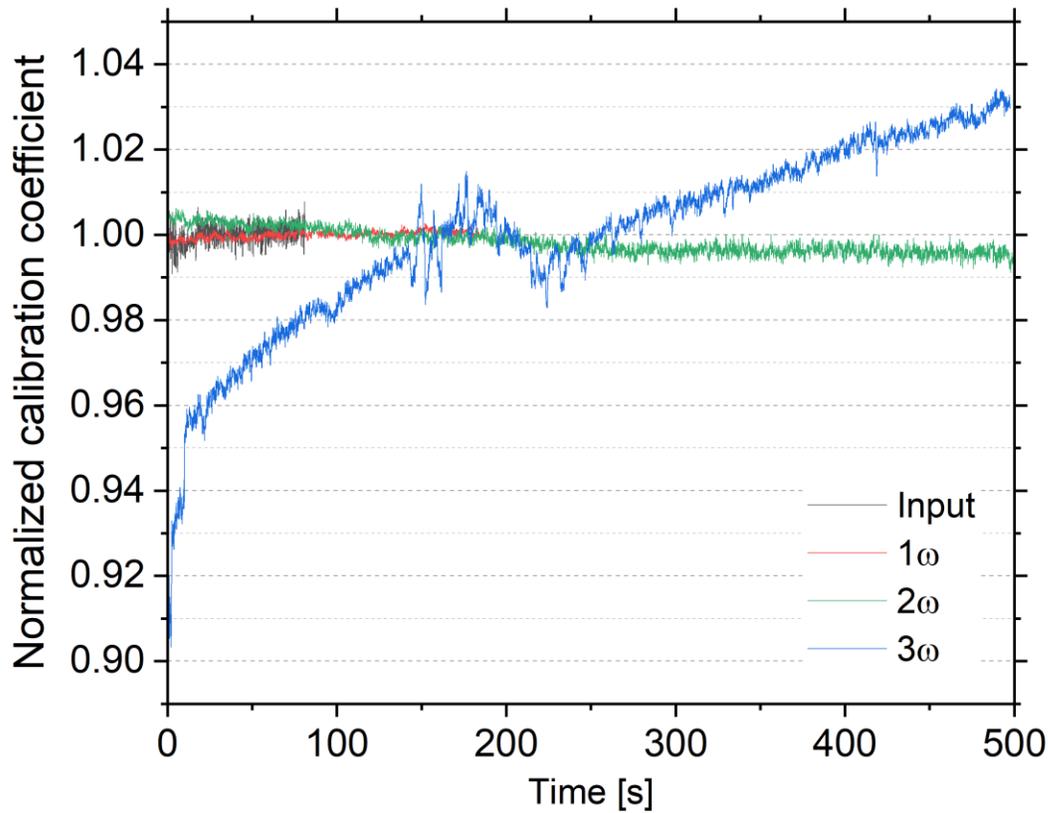

**Figure 2** Temporal evolution of energy meter's calibration coefficients, while calibration coefficient for input energy (1030 nm) and unconverted fundamental *1ω* (1030 nm) stabilized in order of tens of seconds, the second harmonic *2ω* (515 nm) took hundreds of seconds and the third harmonic *3ω* (343 nm) didn't stabilize at all.

Detailed description of the Bivoj laser can be found in [13]. Bivoj laser generated pulses with flat-top temporal profile with duration of 10 ns. The profile can be adjusted arbitrarily and optimized with a closed loop within 14 ns shaping window [14]. Such optimization can be done also for temporal profiles of the frequency-converted beams if needed. The spectral bandwidth of the pulses was around 200 MHz and was given by the temporal shaper as the oscillator operates with 70 kHz spectral bandwidth. However, it was demonstrated that Yb:YAG lasers cryogenically cooled to 77 K can generate pulses with spectral bandwidth of 0.43 nm [15], so seeding Bivoj with broader spectrum is also possible. By optimizing the input and output polarization of the Bivoj system [10], we increased the polarization uniformity, so that around 96% of the energy was in polarization suitable for frequency conversion. The *s* and *p* polarization components of Bivoj beam profile at the input to frequency conversion setup are shown in Figure 3. These results were obtained for the output energy of 92 J, the cooling helium gas flow rate in the main amplifier head was 150 gps and its temperature 120 K.

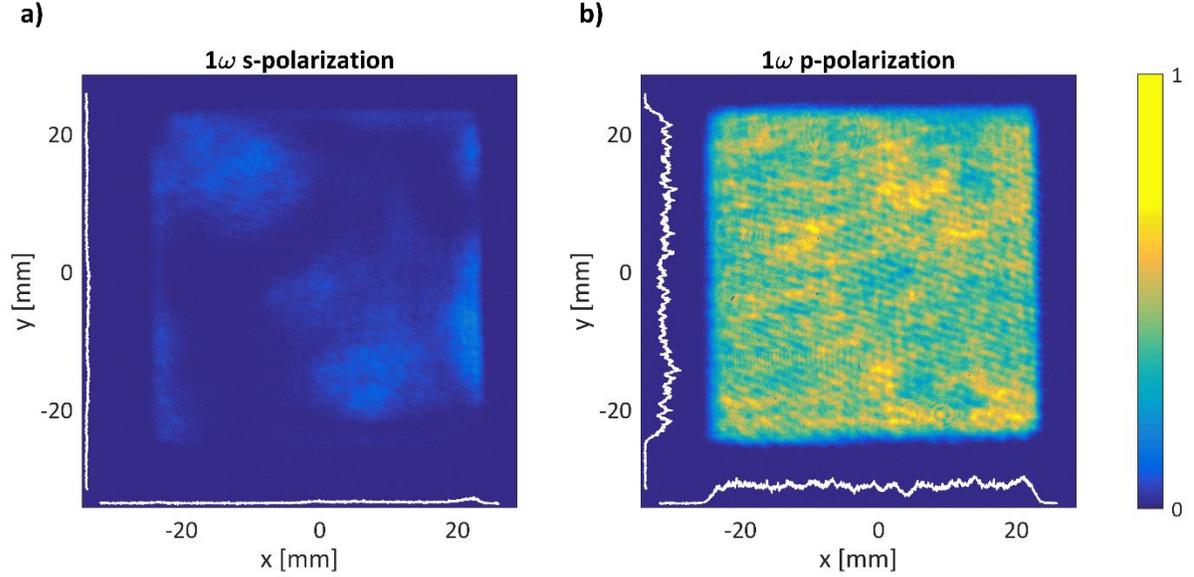

**Figure 3** *s*- and *p*- component profiles of the laser beam after passage through the demagnifying telescope with optimized polarization at the input and output of the laser system. Images were taken after polarizer transmitting vertical polarization with 4% of total energy (a) or horizonal polarization with 96% of total energy (b). Beam profiles at the complementary polarizations were taken under the same conditions and were normalized to sum of both intensities. White lines in the pictures correspond to cross-sections through the center of the beam.

After polarization optimization, the two LBO crystals were inserted into the beam path when using a low energy beam. The input energy on the SHG LBO crystal was then sequentially increased up to 86.5 J and the phase matching angles of both crystals were optimized to maximize the THG yield. The energy conversion data are shown in Figure 4. The energy of THG at 343 nm reached the value of 55 J and conversion efficiency of 63.5%. Conversion efficiency was calculated as ratio between energy exiting THG LBO crystal at third harmonic frequency ($E_{3\omega}$) and input energy at fundamental frequency onto the SHG LBO crystal ($E_{input}$). If we counted only the input energy in the p-polarization, that can be converted to second harmonic by the given LBO crystal (i.e., neglect the energy in the unusable polarization), the conversion efficiency would reach 66%. The ratio would change to count in the input energy in p-polarization only, i.e. $E_{3\omega}/E_{input\text{-}p\text{-}pol}$.

As shown in the previous work [11] the SHG oven was able to stabilize the temperature of the SHG LBO crystal well, when only the second harmonic was generated. However, after the insertion of THG crystal and optimization of its angle, the SHG conversion efficiency started to deteriorate. We were able to recover most of the SHG and THG performance by adjusting the phase matching angle of the SHG crystal only. Adjustments of the THG crystal angle had no effect on the resulting THG performance, if the THG crystal was misaligned and no third harmonic was generated, the feedback disappeared. The back-reflected third harmonic beam was hitting frame of the crystal holder and the crystal itself. Based on this observation, we believe that there was parasitic feedback of the 343 nm radiation that caused rapid heating of the SHG crystal and instability of its temperature stabilization control. While the overall temperature increase of the SHG crystal can be compensated by adjusting its phase matching angle, the induced temperature gradients could not be compensated in such way.

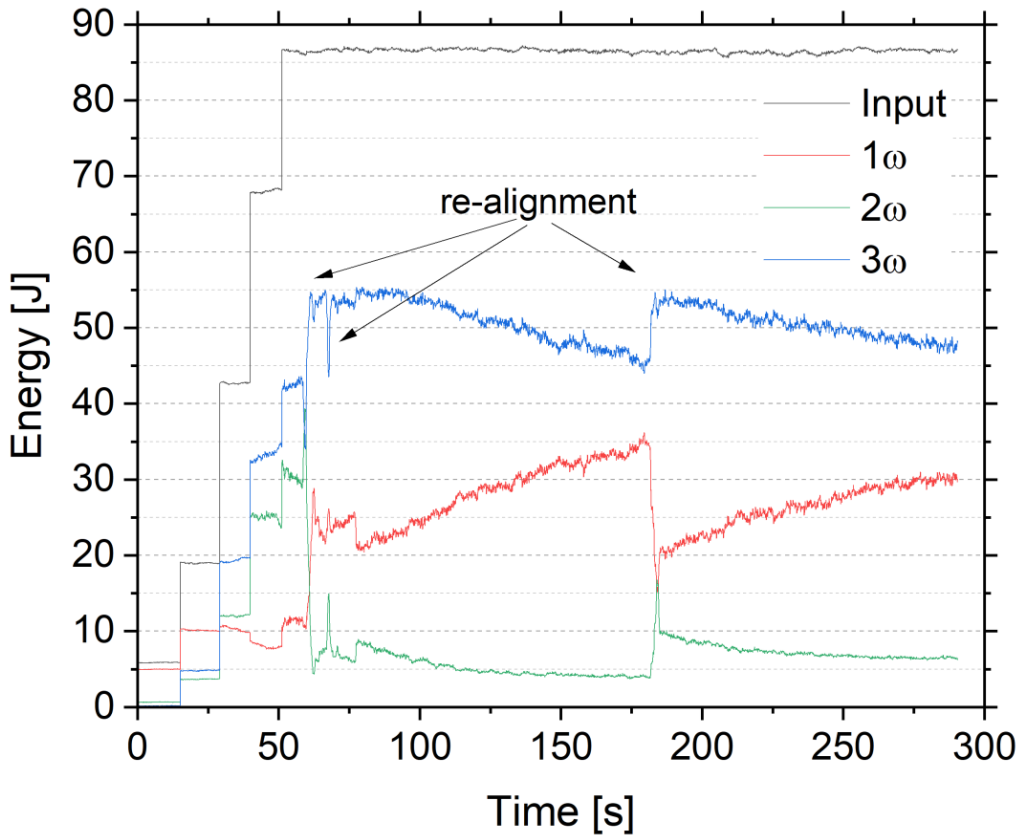

Figure **4** Temporal evolution of the energy of the third harmonic frequency *3ω* together with input energy (1030 nm) and unconverted residual energy on fundamental *1ω* (1030 nm) and second harmonic *2ω* (515 nm) frequencies. Points where crystal phase matching angle was optimized are marked with arrows.

The SHG oven took more than 40 minutes to stabilize the temperature and it required continuous tuning of the phase matching angles to keep the heat sources at constant level. Still, the temperature stability as well as consequent energy stability was achieved in the end as is shown in Figure 5. However, the stable output energy at 343 nm dropped to 46-48 J.

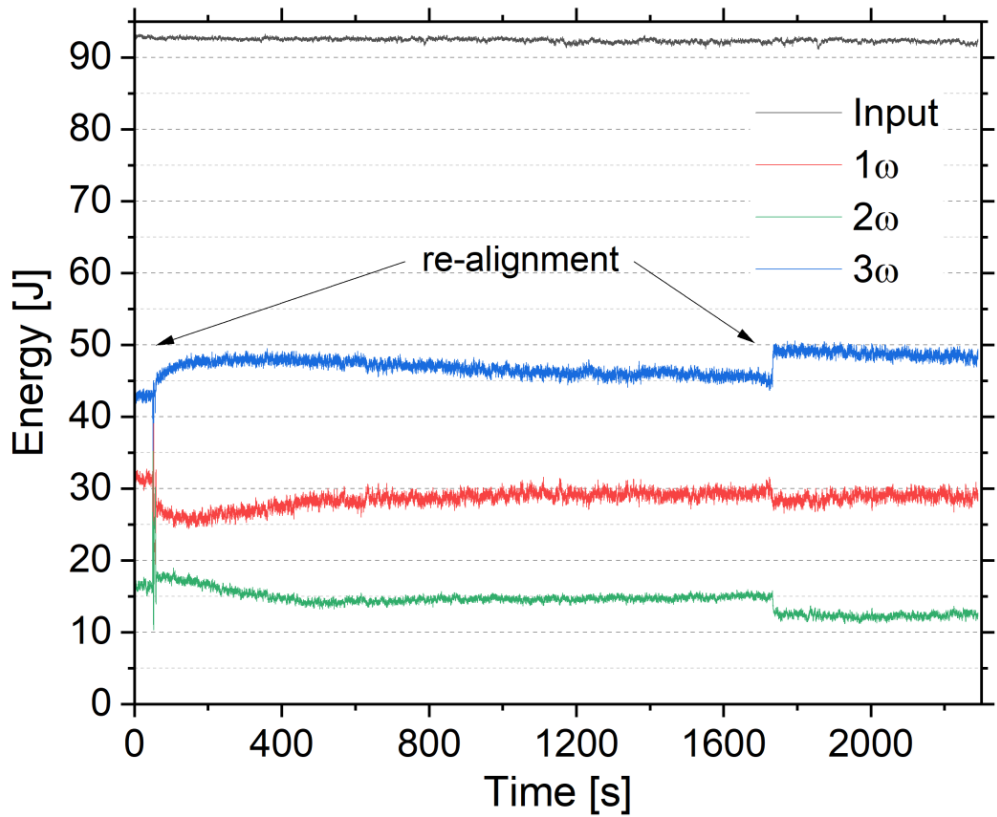

**Figure 5** Temporal evolution of the energy of the third harmonic frequency *3ω* (343 nm) together with input energy (1030 nm) and unconverted residual energy on fundamental *1ω* (1030 nm) and second harmonic *2ω* (515 nm) frequencies after SHG oven temperature stabilization.

Final phase matching angles optimization in Figure 5 (around time of 1700 s) increased output energy to almost 50 J with converison efficiency of 53.5%. If only the convertible energy on p-polarization is taken into account, the conversion efficiency will increase to 55.5%. This final section is shown in Figure 6 in detail.

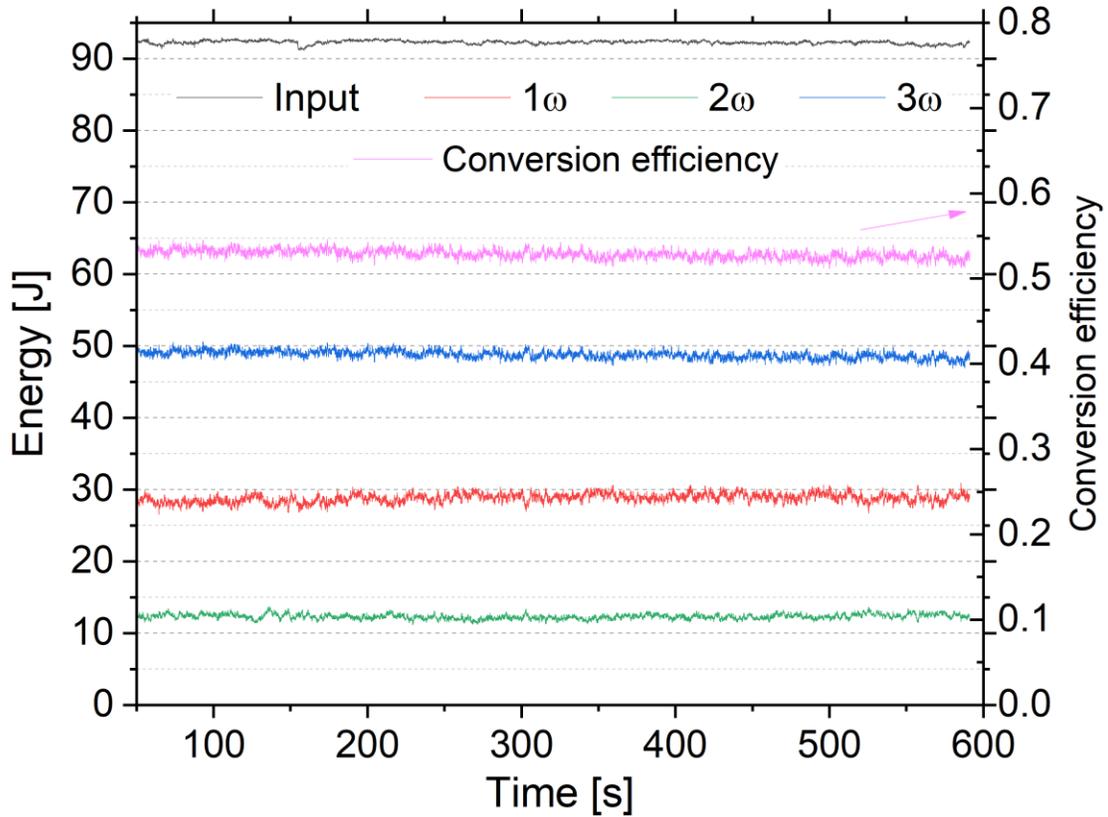

**Figure 6** Temporal evolution of the energy of the third harmonic frequency *3ω* (343 nm) together with input energy (1030 nm) and unconverted residual energy on fundamental *1ω* (1030 nm) and second harmonic *2ω* (515 nm) frequencies after SHG oven temperature stabilization and fine tuning of the SHG LBO phase matching angle. Conversion efficiency $E_{3\omega}/E_{input}$ is shown in pink and is related to the scale on the right.

The energy stability of the input 1030 nm beam was 0.3% RMS and 2.0% P-t-P (peak-to-peak). The energy stability of the 343 nm beam was showing a trend during the experiment as in the beginning it was rather high at around 6% RMS and 22% P-t-P (Figure 4, time 75 – 160 s), which was caused presumably by dynamic heating of the SHG LBO crystal. After thermalization of the SHG crystal and adjustment of the SHG phase matching angle, the energy stability value settled at 1.1% RMS and 7.5% P-t-P (Figure 6).

The near field beam profile at 343 nm corresponding to energy of 55 J at the beginning of the experiment and 50 J obtained after 90 minutes are shown Figure 7. They show that temperature gradient was created inside the crystals and conversion efficiency varied over the crystal cross-section.

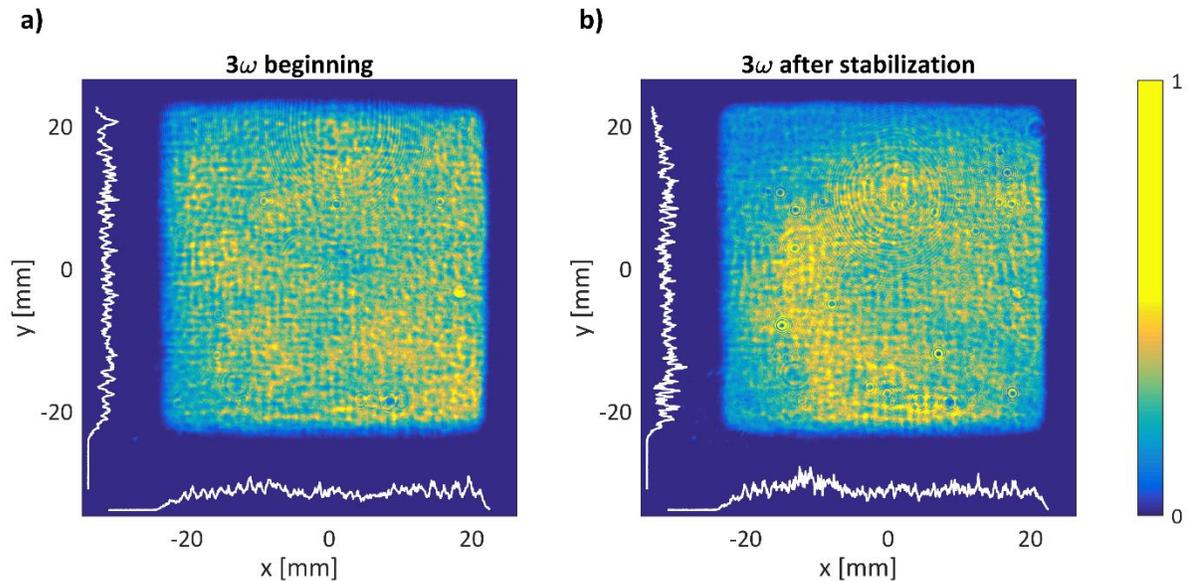

**Figure 7** Beam profile on third harmonic frequency (343 nm) with energy of >50 J at the repetition rate of 10 Hz in the beginning of the experiment (a) and after oven temperature stabilization after 90 minutes (b). The color bar was adjusted to better show the intensity variation in the beam in the presence of hot spots that affected the normalization.

## III. Conclusion

For the first time a high energy THG output of 50 J at 10 Hz was achieved, which presents 4 times increase in terms of average power to the state-of-the-art. The third harmonic generation was realized by frequency doubling and consequent frequency summing of the residual fundamental and frequency doubled beams. This was done using two LBO crystals in temperature-controlled holders. The output beam profile is square super-gaussian with satisfactory energy uniformity. Long thermalization was required due to strong optical feedback between second and third harmonic generation crystals caused by strong reflection from AR coatings. This will be addressed in the continuation of this work. However, after the thermalization, the energy stability of THG output was satisfactory and reached 1.1% RMS.

## Acknowledgement

European Regional Development Fund and the state budget of the Czech Republic project HiLASE CoE (CZ.02.1.01/0.0/0.0/15_006/0000674), project LasApp (CZ.02.01.01/00/22_008/0004573) and Horizon 2020 Framework Programme (H2020) (739573).

## Conflict of Interest

The authors have no conflicts to disclose.

## Data Availability Statement

The data that support the findings of this study are openly available in Zenodo repository at https://doi.org/10.5281/zenodo.12569726, reference [16].